\documentclass[epj,twocolumn]{webofc}
\usepackage[varg]{txfonts}   
\wocname{epj}
\woctitle{LHCP 2013} 
\newcommand{\pp}{\rm pp}

\newcommand{\sqrtsNN}{\sqrt{s_{\rm NN}}}

\newcommand{\gev}{\mathrm{GeV}}
\newcommand{\tev}{\mathrm{TeV}}

\newcommand{\ptrans}{p_{\rm T}}

\newcommand{\Dzero}{{\rm D^0}}
\newcommand{\Dstar}{{\rm D^{*+}}}
\newcommand{\Dplus}{{\rm D^+}}
\newcommand{\Ds}{{\rm D_{s}}}

\newcommand{\vtwo}{v_{2}}
\newcommand{\RAA}{R_{\rm AA}}
\newcommand{\RpPb}{R_{\rm pPb}}
\newcommand{\RAAD}{R_{\rm AA}^{\rm D}}
\newcommand{\RAAB}{R_{\rm AA}^{\rm B}}

\newcommand{\Rcp}{R_{\rm CP}}
\newcommand{\jpsi}{{\rm J/}\psi}
\newcommand{\upsOne}{\Upsilon{\rm (1S)}}
\newcommand{\upsTwo}{\Upsilon{\rm (2S)}}
\newcommand{\upsThree}{\Upsilon{\rm (3S)}}

\newcommand{\QQbar}{ Q\bar{Q}}
\newcommand{\ccbar}{c{\bar c}}

\newcommand{\avTAA}{\langle T_{\rm AA} \rangle}

\newcommand{\Npart}{\langle N_{\rm part} \rangle}
\begin{document}
\title{Heavy-flavour and quarkonia in heavy-ion collisions}
%
%

\author{A. Rossi\inst{1}\fnsep\thanks{\email{andrea.rossi@cern.ch}}
on behalf of ALICE, ATLAS and CMS Collaborations
}


\institute{CERN 
}

\abstract{%
  The comparison of heavy-flavour hadron production in proton-proton, proton-Pb and Pb--Pb collisions
  at the LHC offers the opportunity to investigate the properties of the high-density
  colour-deconfined state of strongly-interacting matter (Quark Gluon Plasma, QGP) that is expected to be formed
  in high-energy collisions of heavy nuclei. 
  A review of the main quarkonium and open heavy-flavour results obtained by the ALICE, ATLAS and CMS experiments
  is presented.
}
\maketitle
\section{Introduction}
\label{intro}
The comparison of quarkonium and open heavy-flavour hadron production in proton-proton, proton-Pb and Pb--Pb collisions
at the LHC offers the opportunity to investigate the properties of the high-density
colour-deconfined state of strongly-interacting matter (Quark Gluon Plasma, QGP) that is expected to form
in high-energy collisions of heavy nuclei~\cite{pbmJSNature}. Due to their large mass, charm and beauty quarks are 
created at the initial stage of the collision in hard-scattering processes with high virtuality ($Q^{2}\gtrsim 4 {\rm m}_{c[b]}^{2}$)
involving partons of the incident nuclei. 
They interact with the medium and lose energy 
via both inelastic (medium-induced gluon radiation, or radiative energy loss)~\cite{gyulassy,bdmps} 
and elastic (collisional energy loss)~\cite{thoma} processes. The loss of energy, sensitive 
to the medium energy density and size, is expected to be smaller for heavy quarks than
for light quarks and gluons, due to the smaller colour coupling factor of quarks with respect to gluons, 
and to the `dead-cone effect', which reduces small-angle gluon radiation for heavy quarks with moderate 
energy-over-mass values~\cite{deadcone}. A sensitive observable is the nuclear modification factor,
defined as $\RAA(\ptrans)=\frac{{\rm d}N_{\rm AA}/{\rm d}\ptrans}{\avTAA\times {\rm d}\sigma_{\pp}/{\rm d}\ptrans}$,
where $N_{\rm AA}$ is the 
yield measured in heavy-ion collisions, $\avTAA$ is the average nuclear overlap function calculated with the 
Glauber model~\cite{glauber} in the considered centrality range, and $\sigma_{\pp}$ is the production cross section 
in pp collisions. A similar observable is the central-to-peripheral ratio ($\Rcp$), in which the yield measured
in central collisions is compared to that measured in peripheral ones scaled by the central-to-peripheral ratio of the number
of nucleon-nucleon binary collisions estimated with the Glauber model. In-medium energy loss determines a suppression, $\RAA$<1, of hadrons at 
moderate-to-high transverse momentum ($\ptrans\gtrsim 2~\gev/c$). 
The dependence of the energy loss on the parton nature (quark/gluon) and mass can be investigated by comparing the nuclear modification factors 
of hadrons with charm $(\RAAD)$ and beauty $(\RAAB)$ with that of pions $(\RAA^{\pi})$, mostly originating
from gluon fragmentation. A mass ordering pattern $\RAA^{\pi}(\ptrans)$<$\RAAD(\ptrans)$<$\RAAB(\ptrans)$ 
has been predicted~\cite{deadcone,Armesto:2005iq}. \\
In heavy-ion collisions with non-zero impact parameter the interaction region exhibits 
an azimuthal anisotropy with respect to the reaction plane ($\Psi_{\rm RP}$) defined
by the impact parameter and the beam direction. Collective effects
convert this geometrical anisotropy into an anisotropy in momentum space that
is reflected in the final hadron azimuthal distribution~\cite{vtwoOllitrault}. The effect, sensitive to the degree of thermalization of the system,
can be evaluated by measuring the $2^{\rm nd}$ coefficient 
of the Fourier expansion of the azimuthal distribution, called elliptic
flow ($\vtwo$). The measurement of heavy-flavour particle $\vtwo$
can provide, at low $\ptrans$, fundamental information on the degree of thermalization 
of heavy quarks in the medium. 
At high $\ptrans$, a non-zero $\vtwo$ can originate from the path-length dependence
of energy loss~\cite{bamps,HFurQMD,CollLPMrad,powlang,whdg,TamuRappEtAl}. \\
%
%

Already in 1986, it was suggested that the formation of $\jpsi$ could be suppressed
in a deconfined medium with free colour charges screening, in a Debye-like way, the QCD $\QQbar$ potential~\cite{jpsiSatz}. The 
screening should be more effective in ``melting'' less 
tightly bound states, leading to the expectation of a sequential suppression pattern of charmonium and bottomonium states, with the higher
states melting at smaller Debye temperatures with respect to the $\jpsi$ and $\upsOne$ ground states~\cite{jpsiSequential}. 
The observation
of $\jpsi$ suppression in Pb--Pb collisions at SPS~\cite{jpsiSPS} was interpreted as a signature of deconfinement. However, a similar 
suppression was observed in Au--Au collisions at $\sqrt{s_{\rm NN}}=200~\gev$ at RHIC~\cite{jpsiPHENIX}, in contrast with the expectation
for a higher suppression in the hotter, denser medium produced at RHIC higher colliding energies. The two results
might be reconciled by an additional production of $\jpsi$ arising from coalescence of c and $\bar{\rm c}$ quarks
in the medium. This $\jpsi$ regeneration mechanism was first proposed by the Statistical Hadronization
Model, which assumes deconfinement and thermal equilibrium of the bulk of $\ccbar$ pairs to
produce $\jpsi$ at the phase boundary by statistical hadronization only~\cite{jpsiSHM}. Later, the transport models
proposed a dynamical competition between the $\jpsi$ suppression by the QGP and the regeneration mechanism,
which enables them to describe also the $\ptrans$ dependence of the $\jpsi$ $\RAA$\cite{jpsiZhaoRapp,jpsiLiuQuXu}. These models
have in common the assumption of deconfinement and some degree of charm quark thermalization. At the LHC 
both effects are expected to intensify: a stronger suppression due to the higher energy density and
temperature and a larger charmonium yield from recombination due to the more copious production of charm quarks
at LHC energies than at SPS/RHIC energies. \\

The measurement of nuclear effects requires the understanding of the production cross-sections of open heavy-flavour and quarkonia in pp collisions,
used as a reference. Open heavy-flavour production is well described by pQCD calculations, like FONLL~\cite{fonll}, relying on the factorization 
approach. For quarkonium, the validity of a factorization approach is not trivial, 
due to the interplay of perturbative and non-perturbative effects, the latter
related to the formation of the non-relativistic $\QQbar$ bound state. For a review of the heavy-flavour pp results at the LHC and of their theoretical modelling
see~\cite{Fasel,CacciariThisProc,Woehri}. A $\RAA$ value different from unity can originate also from initial and final state ``cold-nuclear matter'' effects, 
not related to the formation of a deconfined medium. At LHC energies, nuclear shadowing, which 
reduces the parton distribution functions for partons with nucleon momentum fraction $x$ below $10^{-2}$, %
is expected to be the most important for heavy-flavour production. 
Coherent parton energy loss in cold nuclear matter and  nuclear absorption of the "pre-hadron'' $\QQbar$ pair
before hadronization are also expected to affect quarkonium production~\cite{pPbPredictions}. A correct interpretation of heavy-ion results
demands for the measurement of these effects via the analysis of p--Pb data. \\

In these proceedings, a review of the main quarkonium and open heavy-flavour results obtained by the ALICE, ATLAS and CMS experiments
is presented. The focus is on the measurements performed in Pb--Pb collisions. The preliminary 
results on $\jpsi$ production in p--Pb collisions will be discussed as well. %
\begin{figure*}
  \centering
  \includegraphics[width=7cm,clip]{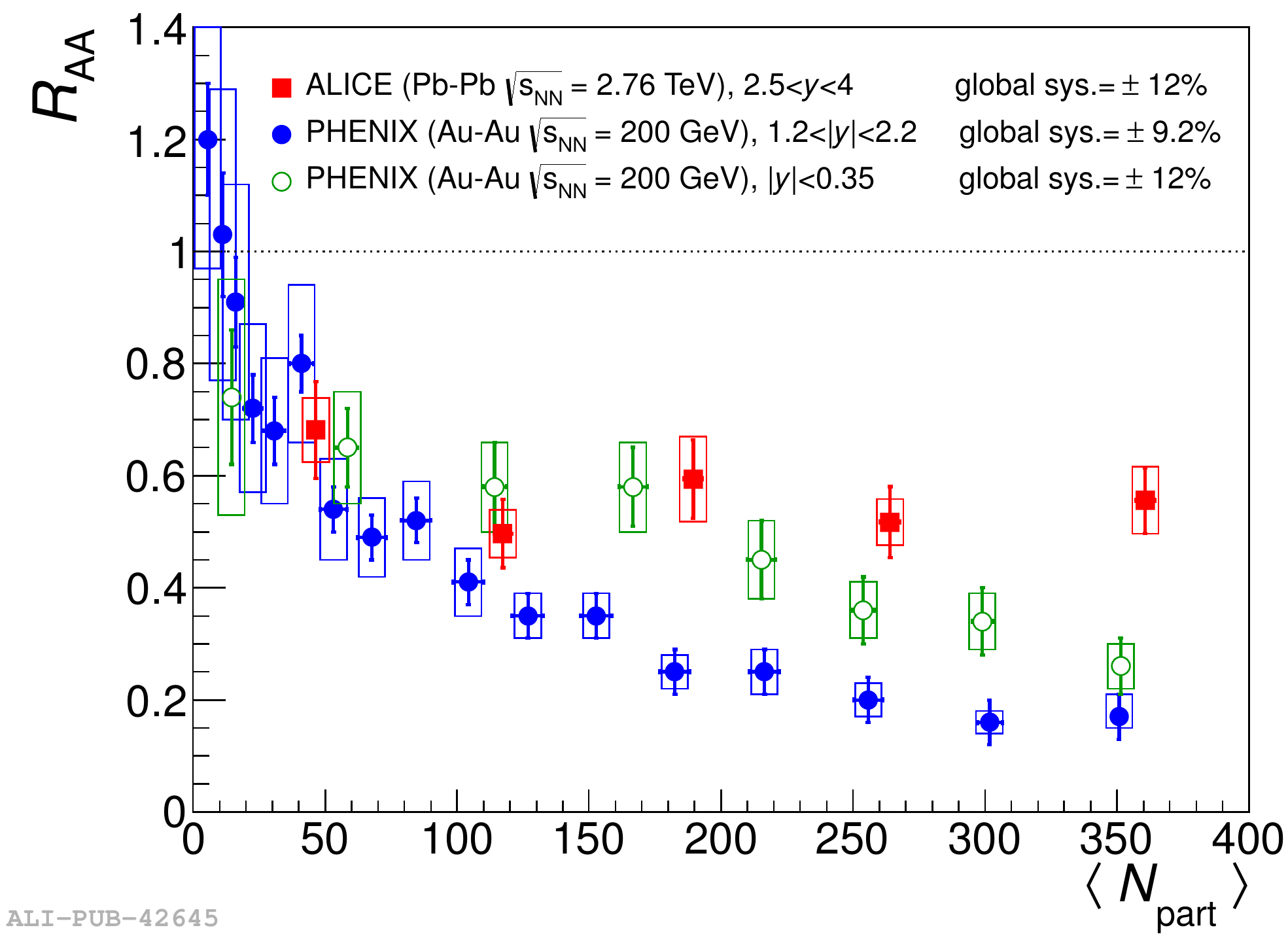}
~~~~~~~~~~~~~~~~~~~~~~  
\includegraphics[width=5.3cm,clip]{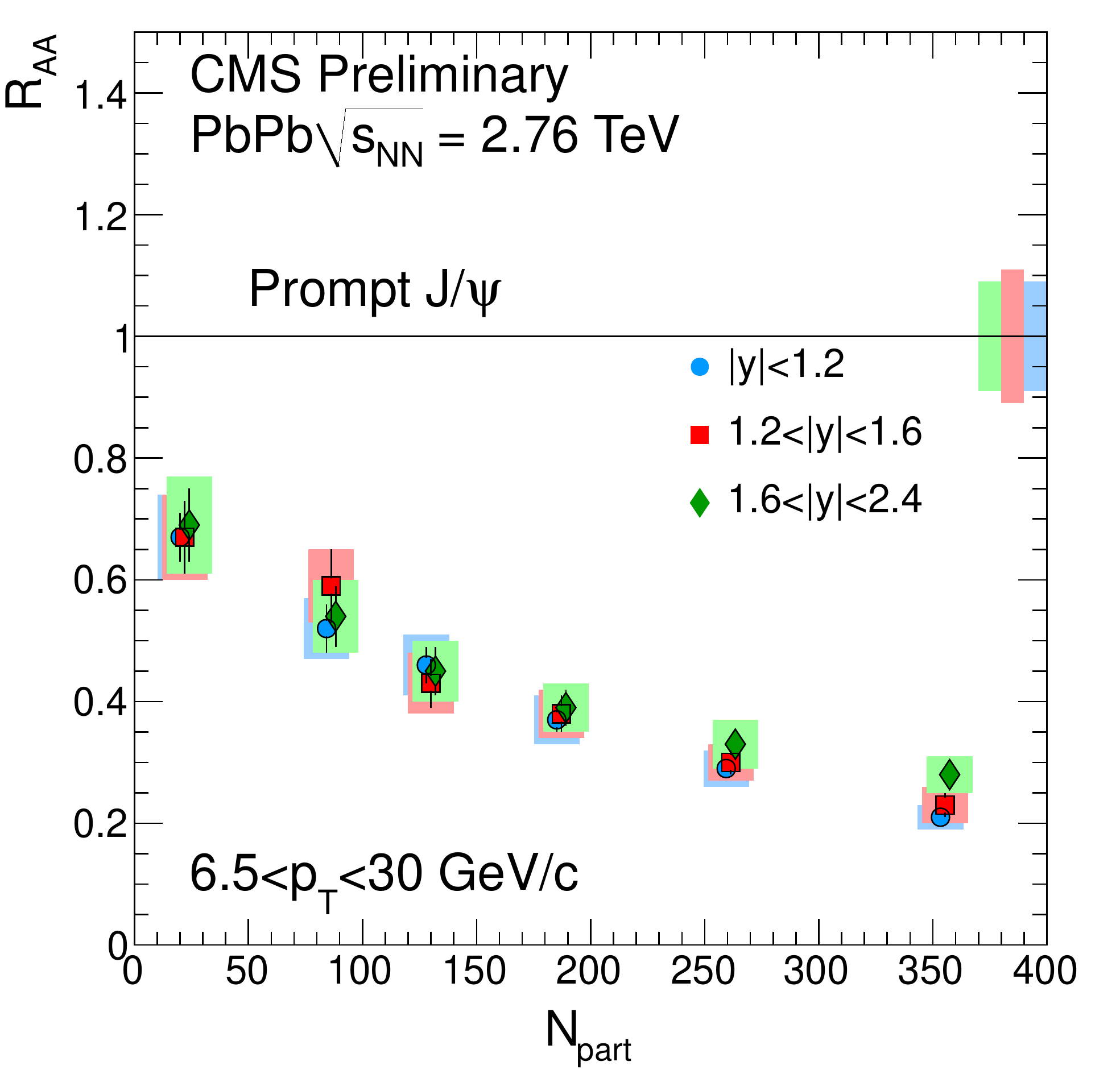}
  \includegraphics[width=7cm,clip]{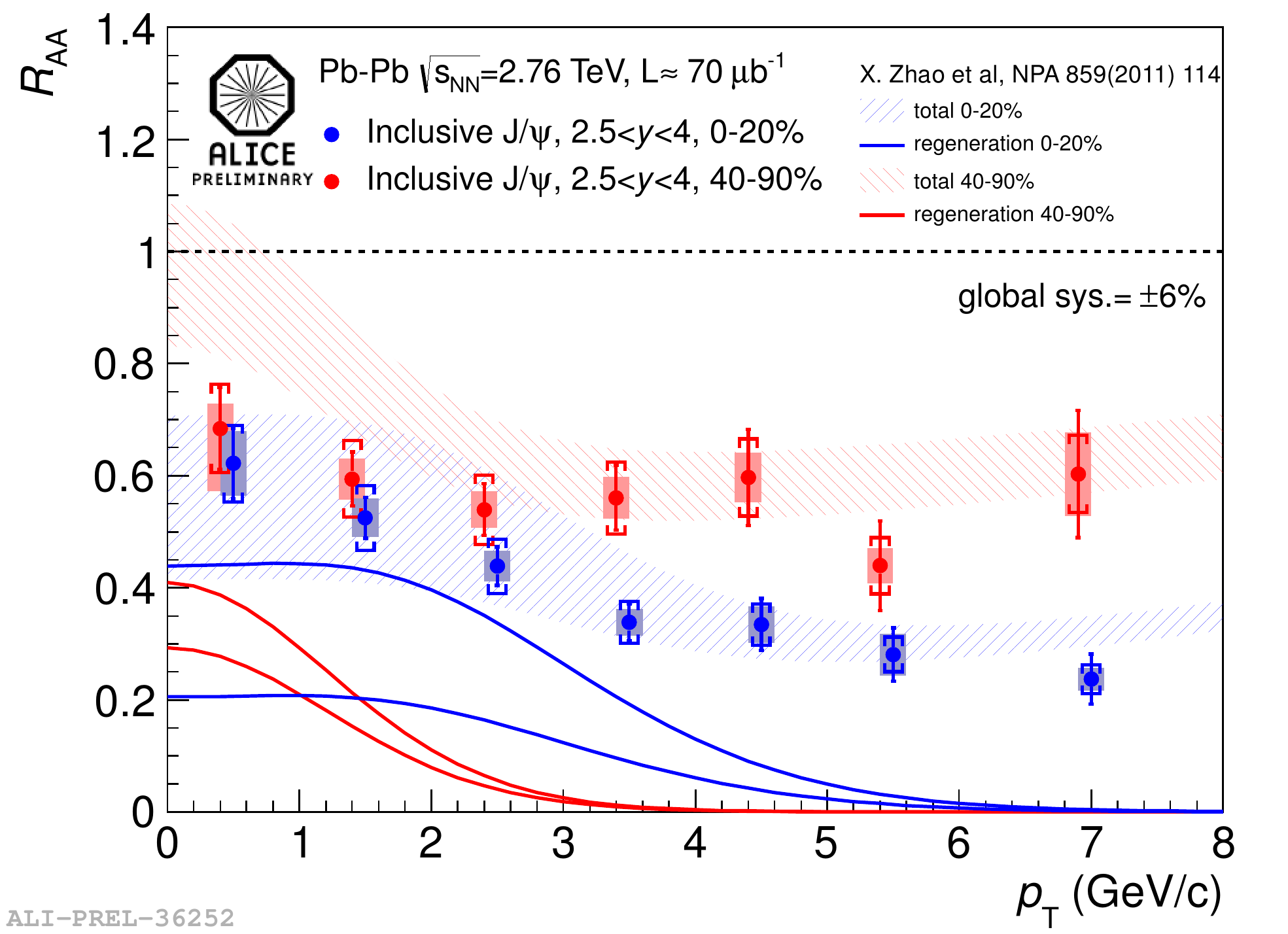}
  \includegraphics[width=7cm,clip]{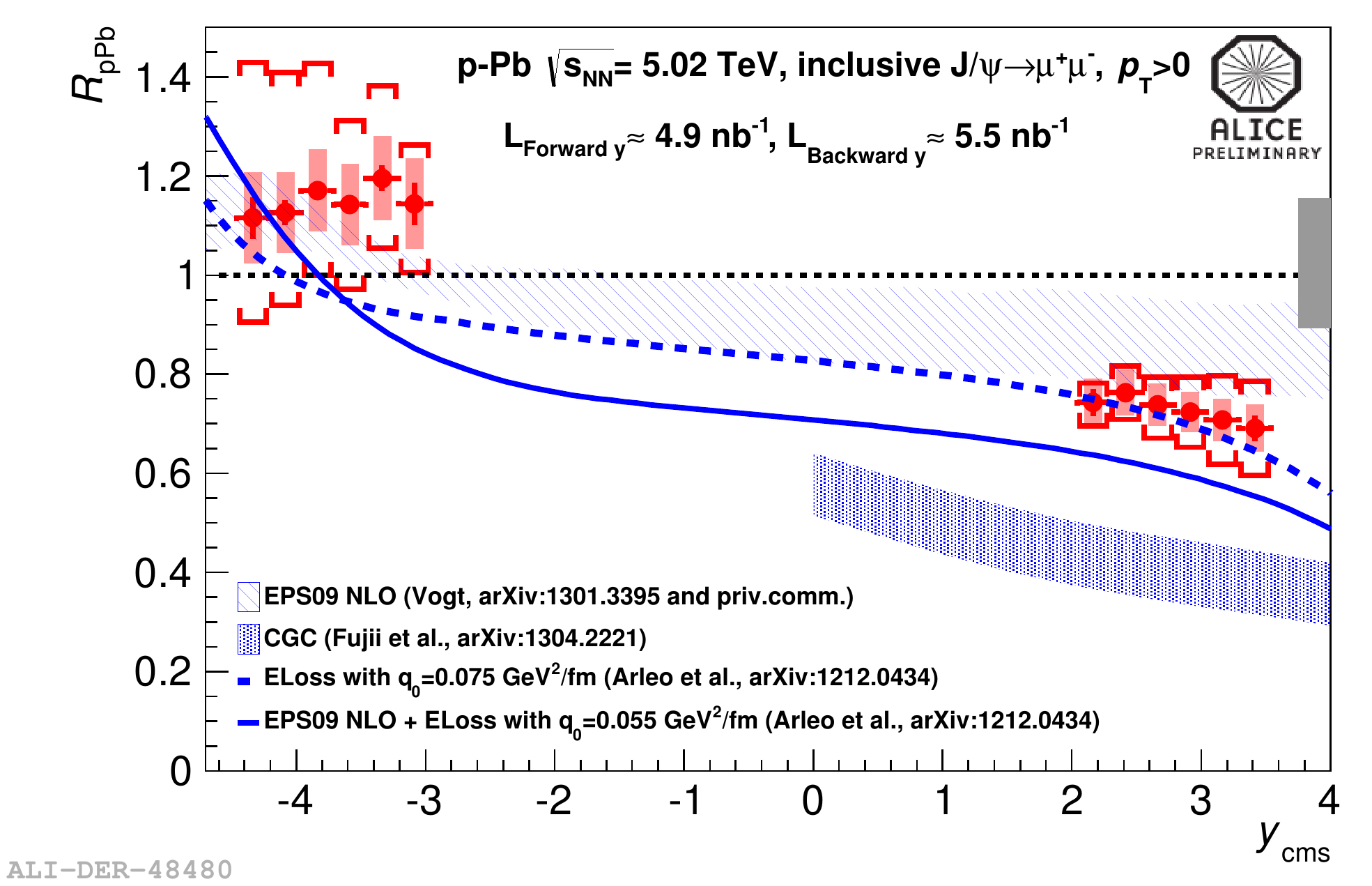}
  \caption{Top: centrality dependence of $\jpsi$ $\RAA$ for $\ptrans$>0 in 2.5<$y$<4 (left, inclusive $\jpsi$, figure from~\cite{ALICEjpsiPaper}) 
    and for 6.5<$\ptrans$<30~$\gev/c$ in different rapidity intervals (right, prompt $\jpsi$, figure from~\cite{CMSjpsiPAS}) 
    measured by ALICE and CMS respectively in Pb--Pb collisions at $\sqrtsNN=2.76~\tev$ at the LHC. ALICE 
    data are compared to the measurements performed by PHENIX in Au--Au collisions at $\sqrtsNN=200~\gev$ at RHIC. Bottom left: $\ptrans$ dependence of $\jpsi$ $\RAA$ 
    compared to the transport model by Zhao~et~al.~\cite{jpsiZhaoRapp}. Bottom right: $\jpsi$ $\RAA$ in p--Pb collisions at $\sqrtsNN=5.02~\tev$ compared to models (see text).}
  \label{figJPsi}       
\end{figure*}
\section{Quarkonia }
\label{sec:quarkonia}
ALICE, ATLAS and CMS measured quarkonium production in Pb--Pb collisions by reconstructing charmonia and bottomonia states in the di-muon decay channel. At mid-rapidity,
ALICE measured $\jpsi$ production in the di-electron channel. Exploiting
the nice complementarity of the kinematic ranges covered by the 3 experiments, quarkonium production is scanned through 4 rapidity units from $\ptrans=0$ to high $\ptrans$. \\

The nuclear modification factor of inclusive $\jpsi$ for $\ptrans$>0 in $2.5$<$y$<$4$ measured by ALICE as a function of the collision 
centrality, expressed in terms of number of nucleons participating to the collisions~\cite{ALICEjpsiPaper}, is compared
in the top left panel of figure~\ref{figJPsi} with the measurements performed by the PHENIX experiment in Au--Au collisions at $\sqrt{s_{\rm NN}}$=200~$\gev$
at RHIC in $|y|$<0.35 and in 1.2<$|y|$<2.2.  At the LHC the suppression is independent from the collision centrality for $\Npart$>100 and a smaller
suppression is observed for most central collisions with respect to RHIC. Both transport models~\cite{jpsiZhaoRapp,jpsiLiuQuXu,jpsiFerreiro} and the statistical
model~\cite{jpsiSHM,jpsiAndronic}, including a relevant contribution from recombination, can reproduce the data reasonably well~\cite{ALICEjpsiPaper,Manceau}. 
The same considerations hold at mid-rapidity, where $\jpsi$ production for $\ptrans$>0 was measured by ALICE 
in the di-electron channel in $|y|$<0.9~\cite{Manceau}. Within the rapidity range of the reported measurement, $\RAA$ decreases by about $40\%$ towards 
forward rapidity. As a function of transverse momentum, $\RAA$ decreases significantly in central collisions, from about 0.65 
at $\ptrans$$\sim$0.5~$\gev/c$ to about 0.2 at $\ptrans$$\sim$7~$\gev/c$, as shown
in the bottom-left panel of figure~\ref{figJPsi}. The trend is well reproduced by the 
transport model by Zhao~et~al.~\cite{jpsiZhaoRapp}, which foresees a substantial contribution from  
regenerated $\jpsi$ at low $\ptrans$, vanishing towards higher $\ptrans$. For $\ptrans$$\gtrsim$6~$\gev/c$, where coalescence should play a minor, if not negligible, role, 
the suppression, measured by ALICE and ATLAS~\cite{ATLASjpsiRcp} for inclusive $\jpsi$ (including $\jpsi$ from b-hadron decays) 
and by CMS for prompt only $\jpsi$~\cite{CMSjpsiPaper,CMSjpsiPAS}, increases 
towards most central collisions (top-right panel of figure~\ref{figJPsi}), does not show a significant dependence on $\ptrans$ and is independent from rapidity. Though the
different trends observed at low $\ptrans$ and high $\ptrans$ are in agreement with the expectations from a scenario in which Debye screening
is overcome by $\jpsi$ production from coalescence at low $\ptrans$, the measurement
of $\RAA$ of higher charmonia states, of $\jpsi$ $\rm{v}_{2}$ and of cold nuclear matter effects is 
required for a conclusive interpretation of the results. Regarding the latter, figure~\ref{figJPsi}
shows on the bottom-right panel the first preliminary measurement of $\jpsi$ nuclear modification factor in p--Pb collisions ($\RpPb$), reported by ALICE. In the figure,
negative (positive) values of rapidity refer to the case in which the Pb(p) beam moves towards the ALICE muon
spectrometer used for the measurement. The EPS09 NLO shadowing calculations~\cite{epsZeroNine,pPbshadowingVogt} and 
models including coherent parton energy loss~\cite{jpsipPbArleo} can reproduce
the data, while the specific Colour Glass Condensate model described in~\cite{cgcFuji} underestimates the measurement. Calculations based on shadowing EPS09 
are not able to reproduce the Pb--Pb measurement~\cite{ALICEjpsiPaper}, indicating that the suppression observed in this latter system is a hot medium effect. $\jpsi$ 
coming from $\ccbar$ recombination inherit the azimuthal anisotropy of charm quarks and, therefore, should show a non-zero $\vtwo$, if charm quarks take part
in the collective expansion of the system. %
%
\begin{figure}
  \centering
~~~
  \includegraphics[width=7cm,clip]{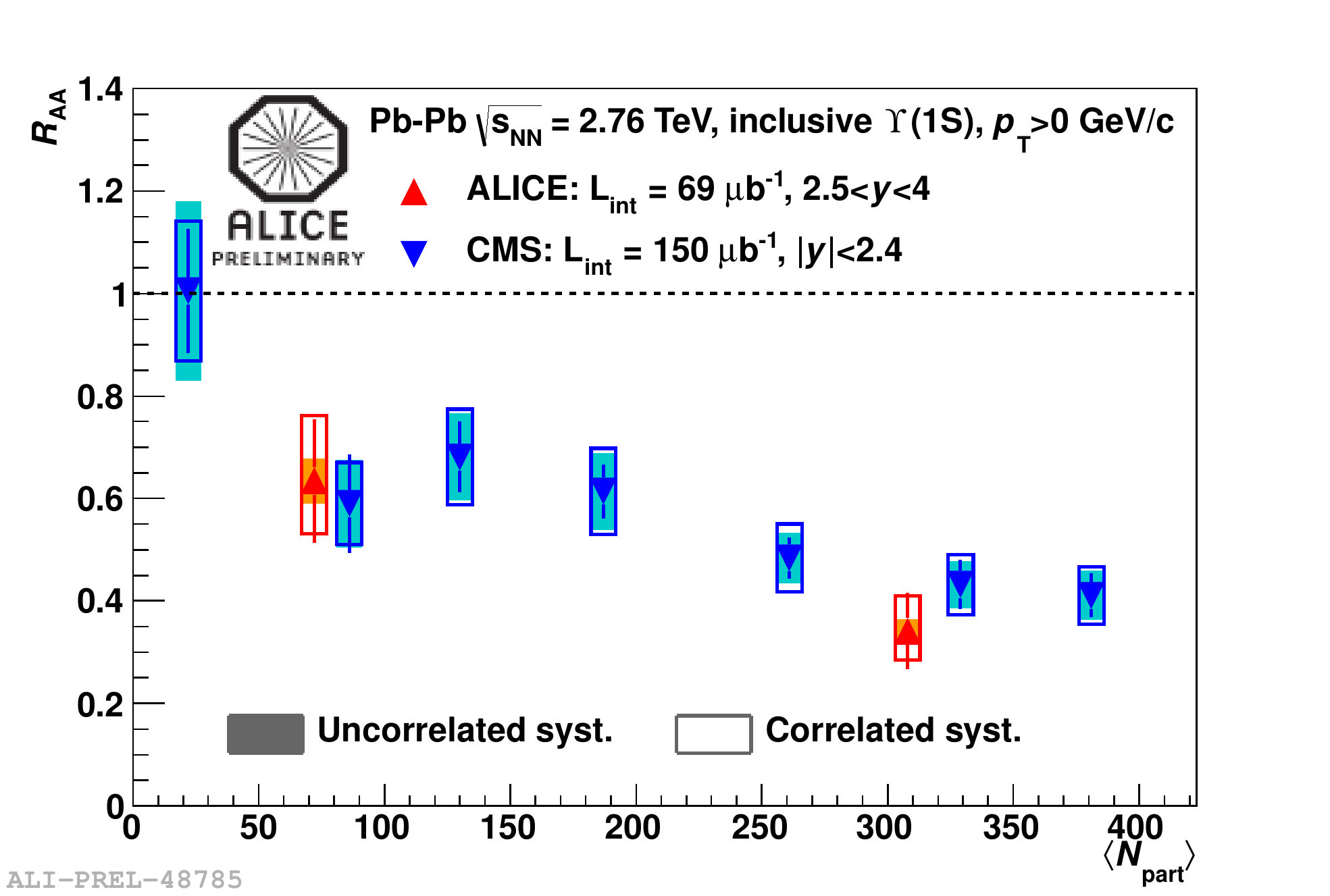}
  \includegraphics[width=6.5cm,clip]{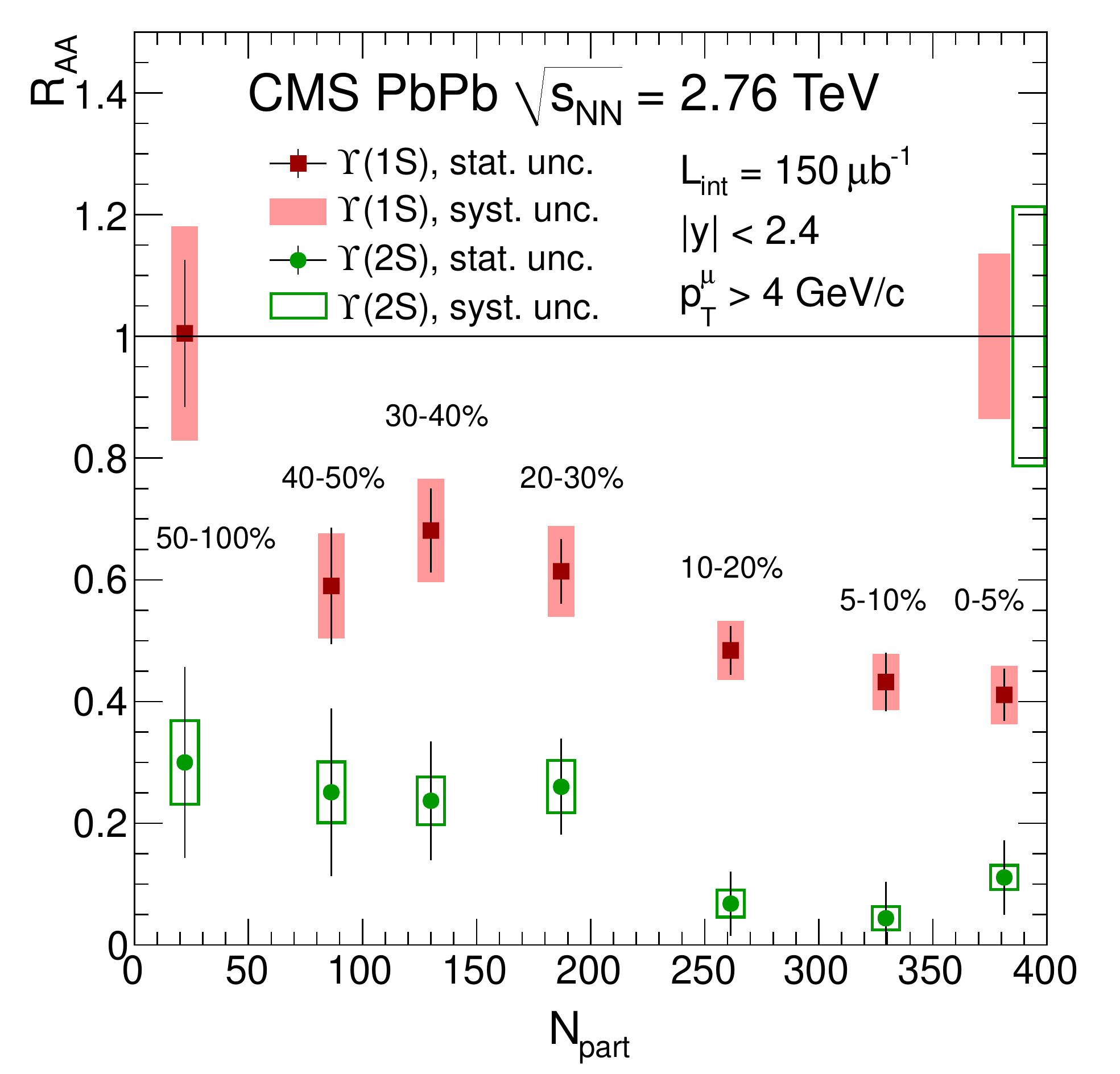}
  \caption{Top: $\upsOne$ $\RAA$ measured by ALICE in 2.5<$y$<4~\cite{Manceau} and by CMS in $|y|$<2.4~\cite{CMSupsilon} as a function of centrality. Bottom: comparison
    of $\upsOne$ and $\upsTwo$ $\RAA$ measured by CMS~\cite{CMSupsilon}.}
  \label{fig:bottomonia}       
\end{figure}
%
A hint of this was observed by ALICE~\cite{jpsiVtwo}, but the statistical
uncertainties are still too large and prevent a conclusion. Concerning
heavier charmonia states, a preliminary hint for a double ratio of $\psi(2S)$ to $\jpsi$ in Pb--Pb and pp larger than unity 
was released by CMS~\cite{psiTwosCMS} for 3<$\ptrans$<30~$\gev/c$ and 1.6<$|y|$<2.4, contrasting 
the expectation of a higher suppression for higher states. The
measurement is however affected by large uncertainties, mainly related to the pp reference, and the preliminary measurement 
by ALICE~\cite{jpsiPsiTwosArnaldiQM} for $\ptrans$>3~$\gev/c$ and 2.5<$y$<4 does not substantiate it. At the moment, there is no
measurement of $\chi_{c}$ $\RAA$ available. \\

The much smaller abundance of bottom quarks with respect to charm quarks should contain
the potential contribution of coalescence to bottomonia production to a negligible level. The $\upsOne$ $\RAA$ 
shown in the top panel of figure~\ref{fig:bottomonia}, measured for $\ptrans$>0
by CMS at mid-rapidity ($|y|$<2.4) and by ALICE at forward rapidity (2.5<$y$<4), decreases from
about 1 in peripheral events to about 0.4 in central events. No significant dependence on 
rapidity can be appreciated within the statistical and systematic uncertainties~\cite{Manceau}. The observed
$\RAA$ is influenced by the suppression level of higher states, whose decay yields $\sim$50\% of 
the $\upsOne$ production in pp collisions at high $\ptrans$. CMS measured the centrality dependence of $\upsTwo$
$\RAA$, which is compared, in the bottom panel of figure~\ref{fig:bottomonia}, to the ground state $\RAA$~\cite{CMSupsilon}. $\RAA$ 
goes from about $0.3$ in peripheral collisions to $\lesssim 0.1$ in central collisions. Thus,
a much stronger suppression is observed with respect to the ground state, in qualitative agreement with the 
expectation of a sequentially increasing suppression of higher quarkonium states, less tightly bound and with a smaller Debye 
temperature. This scenario is also supported by the data reported by CMS in~\cite{CMSupsilon} 
on the $\upsOne$, $\upsTwo$ and $\upsThree$ $\RAA$ in minimum bias collisions,
which also suggest that the direct $\upsOne$ is almost unsuppressed. \\

\begin{figure}
  \centering
  \includegraphics[width=7cm,clip]{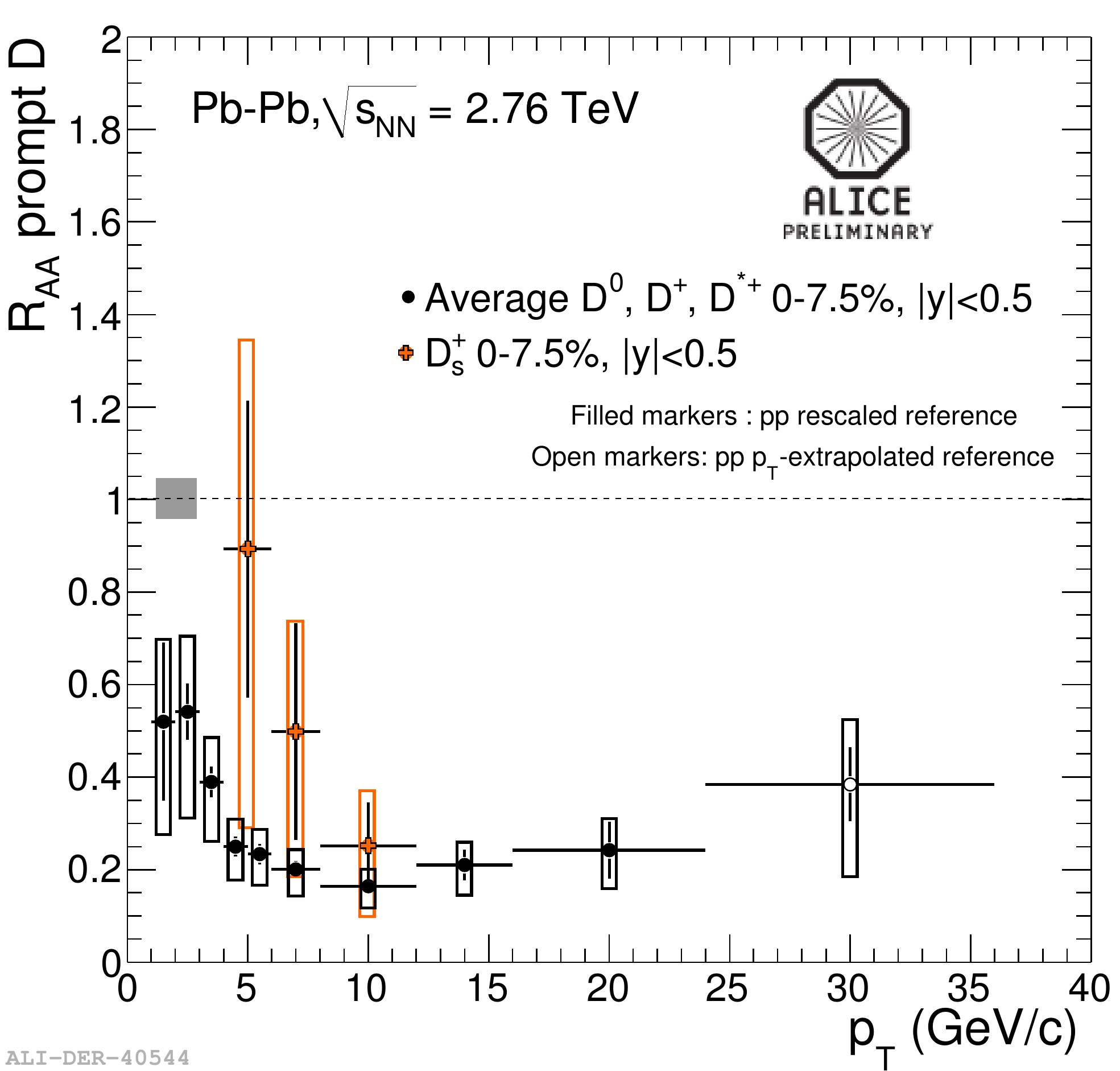}
  \caption{Average of $\Dzero$, $\Dplus$ and $\Dstar$ $\RAA$ and $\Ds$ $\RAA$ as a function of $\ptrans$, measured in 0-7.5\% central Pb--Pb collisions
    by ALICE~\cite{zaidaQM,innocentiDs}.}
  \label{figDraa}       
\end{figure}
%
%
%
%
%
\section{Open heavy-flavour}
\label{sec:openHF}
\begin{figure}
  \centering
  \includegraphics[width=8cm,clip]{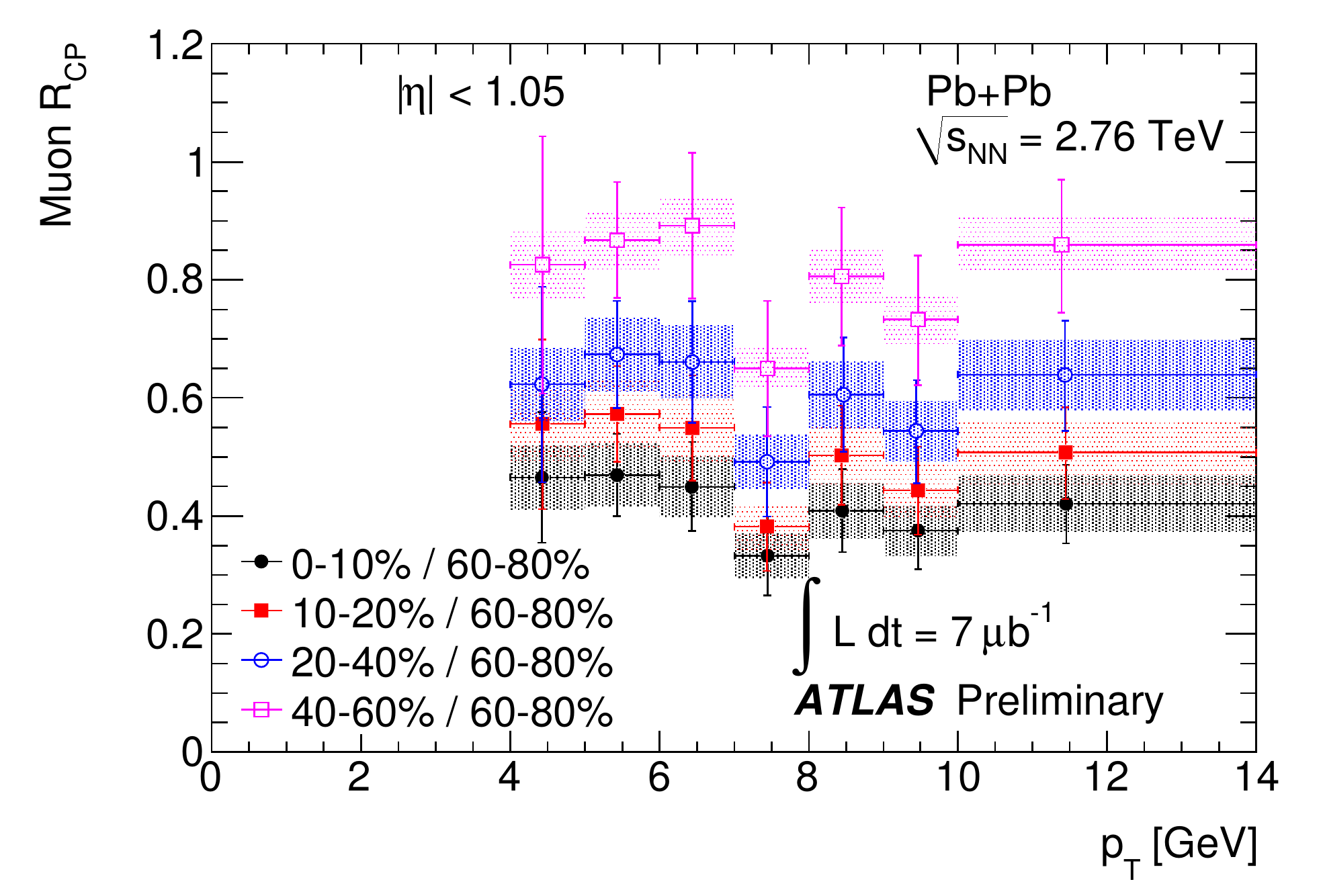}
  \caption{Central-to-peripheral ratio ($\Rcp$) of muons from heavy-flavour hadron decay as a function of $\ptrans$ in different centralities, measured by ATLAS~\cite{ATLASmuonRcp}.}
  \label{figMuonRcp}       
\end{figure}

\begin{figure*}
  \centering
  ~~~
  \includegraphics[width=6.1cm,clip]{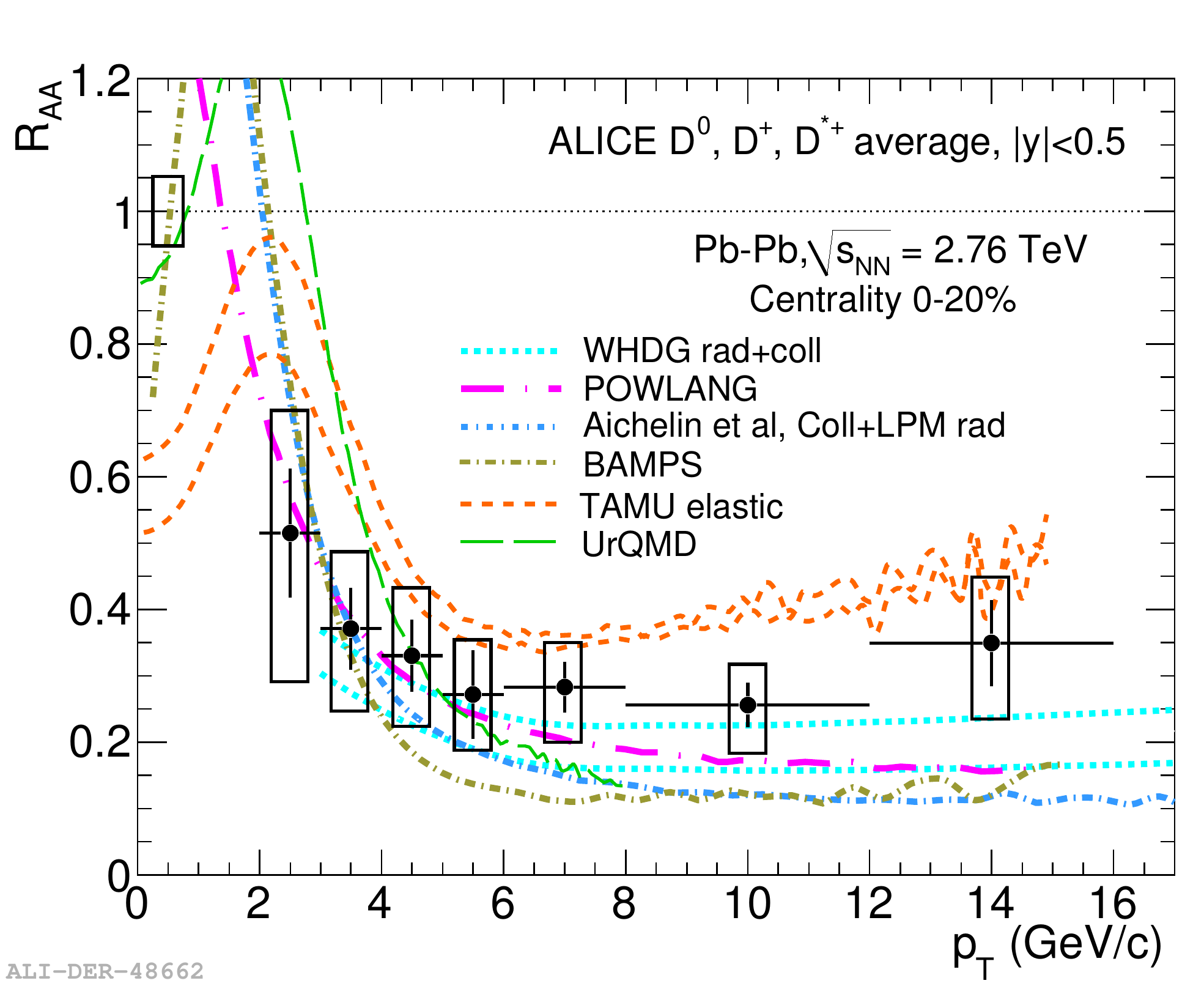}
  \includegraphics[width=6.5cm,clip]{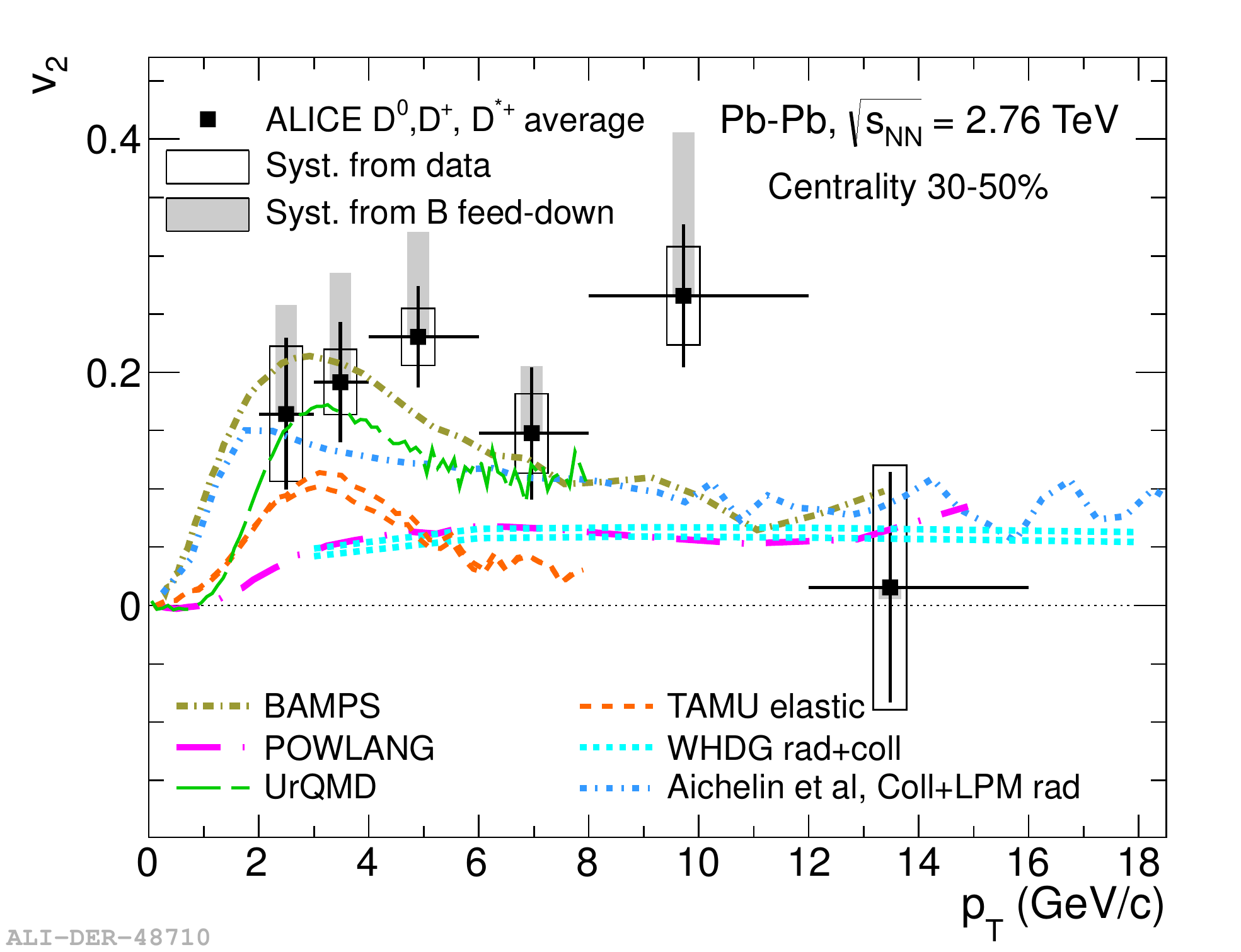}
  \includegraphics[width=12.3cm,clip]{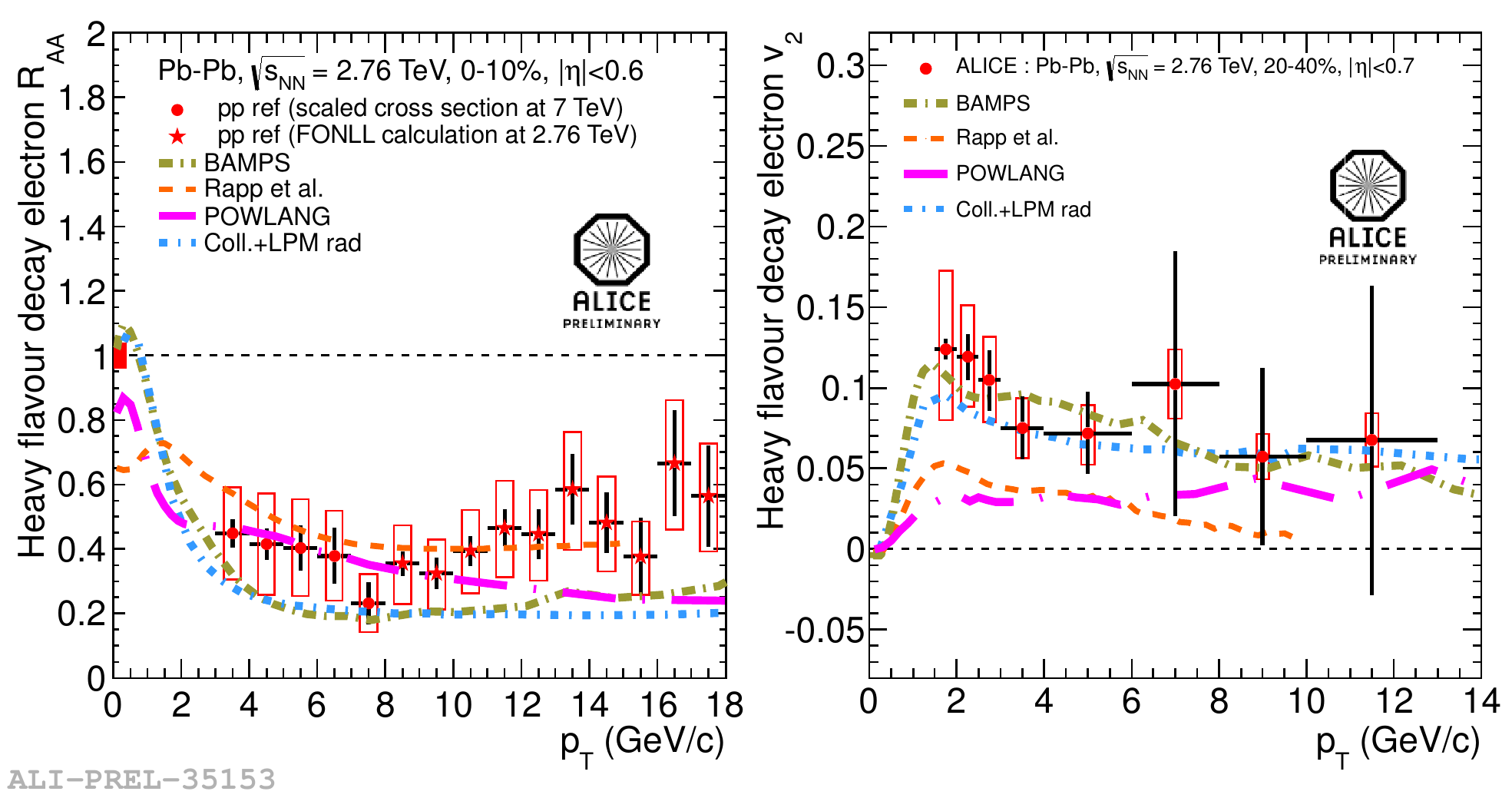}
  \caption{$\RAA$ (left) and $\vtwo$ (right) of D mesons (top) and electrons from heavy-flavour hadron 
    decay (bottom), compared to models: BAMPS~\cite{bamps}, UrQMD~\cite{HFurQMD}, Coll+LPM rad~\cite{CollLPMrad}, POWLANG~\cite{powlang}, 
    WHDG rad+coll~\cite{whdg}, and TAMU elastic (Rapp~et~al.)~\cite{TamuRappEtAl}.}
  \label{fig:DandEleRaaV2models}       
\end{figure*}

Several observables were used to address open charm and beauty production in Pb--Pb collisions: the production of prompt D 
mesons from the full reconstruction of hadronic decay channels at mid-rapidity was measured by ALICE~\cite{aliceDmesons,zaidaQM}, leptons from
semi-leptonic decays of heavy-flavour particles were studied by ALICE (electrons in $|\eta|$<0.6~\cite{Bianchin}, muons in 2.5<$y$<4~\cite{muonPRL}) 
and by ATLAS (muons in $|\eta|$<1.05~\cite{ATLASmuonRcp}), while the production of secondary $\jpsi$ from B decay was measured 
by CMS~\cite{CMSjpsiPAS}. Figure~\ref{figDraa} shows the $\ptrans$ dependence of the average of $\Dzero$, $\Dplus$ and $\Dstar$ meson $\RAA$ 
in 1<$\ptrans$<36~$\gev/c$ in central 0-7.5\% collisions: a strong suppression is observed, by about a factor 5 around $10~\gev/c$, with 
a relevant dependence on $\ptrans$. Within the statistical and systematic uncertainties, the D 
meson $\RAA$ is compatible with that of charged hadrons~\cite{zaidaQM}. The ongoing analysis
of data from the p--Pb run will clarify how much of the observed suppression can be ascribed to a medium effect. However, the main cold nuclear matter
effect is expected to be nuclear shadowing, 
which should yield a relatively small effect for $\ptrans$$\gtrsim$5~$\gev/c$~\cite{aliceDmesons}. Thus, the observed suppression can be considered 
an evidence of in-medium charm quark energy loss. In 
the same figure, the result of the first measurement of the production of the charmed and strange $\Ds$ meson in heavy-ion collisions
is reported~\cite{innocentiDs}. A suppression similar to that of non-strange D mesons is measured in 8<$\ptrans$<12~$\gev/c$. At lower $\ptrans$, the $\Ds$ $\RAA$ is compatible 
with that of the non-strange mesons within the large statistical and systematic uncertainties, but there is an intriguing 
hint for a smaller suppression of $\Ds$ production. A more precise 
measurement, achievable with larger data samples from future LHC runs, might pose quantitative implications for models
which include hadronization via coalescence and foresee an enhancement of $\Ds$ production deriving from the larger
yield of strange quarks in a deconfined medium~\cite{TamuRappEtAl,DsRaphelsky}. In figure~\ref{figMuonRcp}, the 
central-to-peripheral ratio ($\Rcp$) for muons from heavy-flavour hadron decay
measured by ATLAS at mid-rapidity for $\ptrans$>4~$\gev/c$ in different centrality classes is shown. A clear suppression, 
increasing with centrality, is observed, without any appreciable dependence on transverse momentum. In central collisions, a similar
suppression was observed by ALICE for electrons (figure~\ref{fig:DandEleRaaV2models}, bottom left) and muons~\cite{muonPRL} from heavy-flavour hadron decay, 
at central and forward rapidity respectively. In pp collisions, 
according to FONLL predictions~\cite{fonll}, the yield of electrons (muons) from beauty hadron decays exceeds that of electrons (muons) from
charm hadron decays for $\ptrans$$\gtrsim$5~$\gev/c$. Therefore, 
%
the measured $\RAA$ values indicate that the production of beauty hadrons
at high $\ptrans$ is also suppressed in central heavy-ion collisions. A direct evidence of this 
\begin{figure}
  \centering
  \includegraphics[width=7cm,clip]{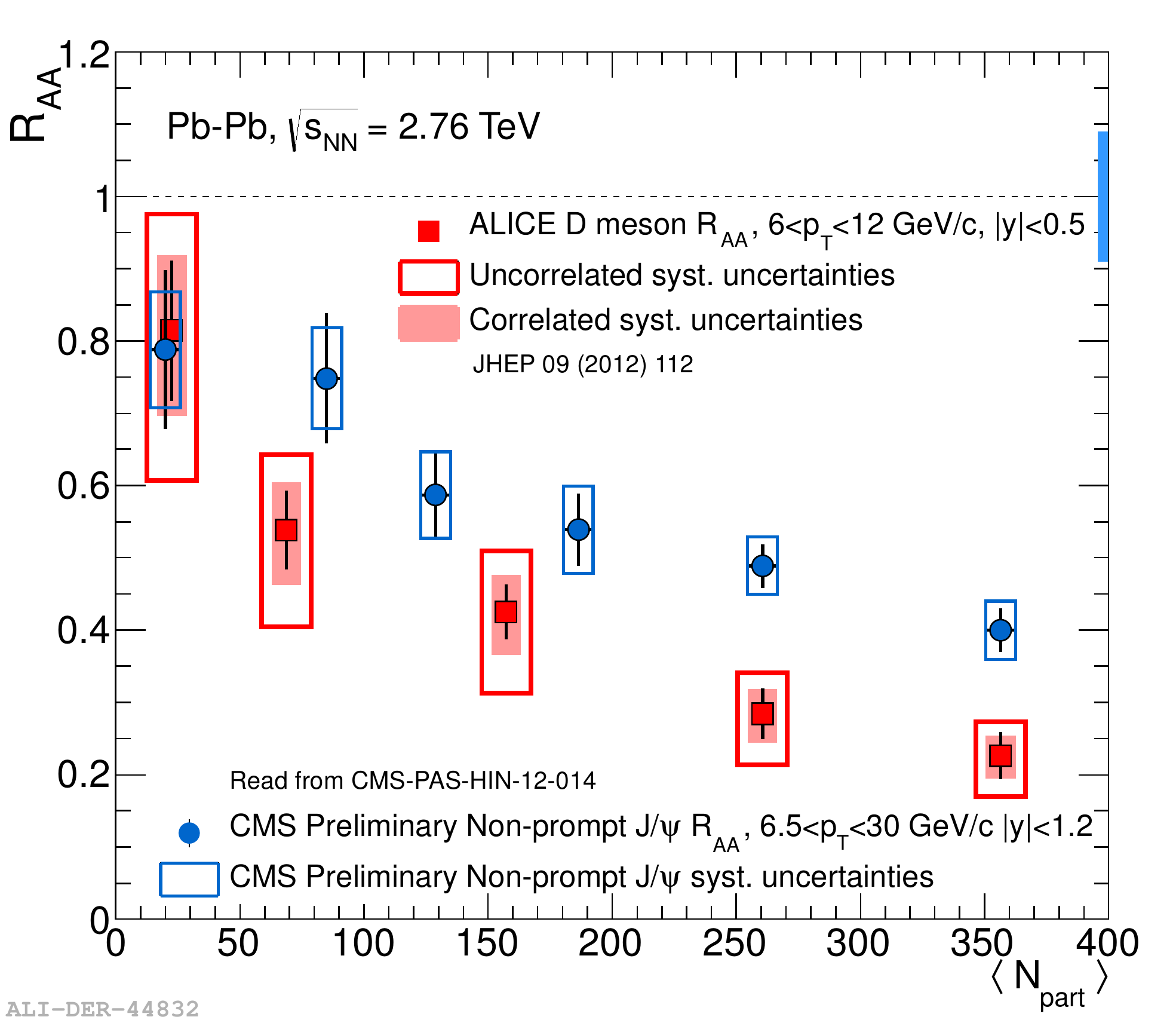} 
  \caption{Comparison of the centrality dependence of the $\RAA$ of D mesons in 6<$\ptrans$<12~$\gev/c$ and secondary $\jpsi$ from
    b-hadron decays in 6.5<$\ptrans$<30~$\gev/c$, measured by ALICE~\cite{aliceDmesons} and CMS~\cite{CMSjpsiPAS} respectively.}
  \label{figAliceDcmsJpsiB}       
\end{figure}
comes from the measurement of secondary $\jpsi$ from beauty-hadron decays performed by CMS~\cite{CMSjpsiPAS}: $\RAA$ decreases
from about 0.8 in peripheral collisions to about 0.4 in central collisions, as shown in figure~\ref{figAliceDcmsJpsiB}
for $\jpsi$ in $|y|$<1.2 and 6.5<$\ptrans$<30~$\gev/c$. The comparison with the centrality dependence
of D meson $\RAA$, displayed in the same figure, highlights a first indication for the mass
dependence of in-medium partonic energy loss, with $\RAAB$>$\RAAD$. However, a firm and quantitative conclusion is prevented
by the fact that the kinematic ranges of the measurements, if considered with respect to the parent quarks, are 
not the same. \\

In the top right panel figure~\ref{fig:DandEleRaaV2models} the first measurement of D meson elliptic flow 
in heavy-ion collisions is shown. The measurement, performed by ALICE in the 30-50\% centrality range~\cite{Dmesonv2}, exploits the event plane method, in which 
the correlation of the particle azimuthal angle ($\phi$) to the reaction plane $\Psi_{\rm RP}$ is analyzed. The 
reaction plane is estimated via the event plane $\Psi_{2}$, by the so-called $Q_{2}$-vector, which is obtained
from the azimuthal distribution of a (sub-)sample of tracks in the event~\cite{VoloshinV2}. Although
the uncertainties are sizable, a significantly non-zero $\vtwo$ is observed in 2<$\ptrans$<6~$\gev/c$, with
an average of the measured values in this range around 0.2. A positive $\vtwo$ is also observed for $\ptrans$>6~$\gev/c$, which most likely originates
from the path-length dependence of the partonic energy loss, although the large uncertainties do not allow
a firm conclusion. The measured D meson $\vtwo$ is comparable in magnitude to that of charged particles,
which is dominated by light-flavour hadrons~\cite{hadronV2}. 
This suggests that low momentum charm quarks take
part in the collective motion of the system. A further indication in this direction is provided by the $\vtwo$ 
of electrons from heavy-flavour decay (same figure, bottom right panel), measured by ALICE with the event plane method
in the 20-40\% centrality range. The measured D meson and electron $\vtwo$ tend to favour models like 
BAMPS~\cite{bamps}, UrQMD~\cite{HFurQMD} and Coll+LPM radiative~\cite{CollLPMrad} that predict a larger 
anisotropy at low $\ptrans$. However, as shown on the left of the same figure, the same models have difficulties in reproducing also the 
D meson and heavy-flavour electron nuclear modification factors in the whole measurement momentum ranges, whereas 
models like POWLANG~\cite{powlang}, WHDG rad+coll~\cite{whdg} and TAMU (Rapp~et~al.)~\cite{TamuRappEtAl} are more 
challenged by the description of heavy-flavour elliptic flow. 
\section{Conclusions}
The smaller $\jpsi$ suppression measured in central heavy-ion collisions at LHC with respect to RHIC and
the observed centrality and $\ptrans$ trends favour a scenario with a significant contribution to $\jpsi$ production 
at low $\ptrans$ from recombination of deconfined $c$ quarks, vanishing at higher $\ptrans$ where
Debye screening determines the production features. The observation of a larger suppression
of $\upsTwo$ and $\upsThree$ than $\upsOne$ supports the Debye-screening picture of a sequential suppression
of states following the hierarchy of binding energies. However, any firm conclusion is still prevented by the lack of information on the 
suppression of higher states, where, especially in the charmonia case, the measurements are characterized by a
poor precision (for $\psi(2s)$) or are absent (for $\chi_{c}$ and $\chi_{b}$), by the theoretical prediction uncertainties, mostly
dominated by the uncertainty on the total charm production cross-section in Pb--Pb collisions, and by the uncertainties related
to cold-nuclear matter effects, the nature of which should be clarified by the ongoing analysis of p--Pb data. More data
from future runs at the LHC should allow for a more precise measurement of $\jpsi~\vtwo$, that, beside 
addressing recombination, could also help to assess the degree of 
charm quark thermalization in the medium. An indication of charm thermalization is provided by the measurement of a non-zero
D meson $\vtwo$ at low $\ptrans$, with values similar to those of charged hadron $\vtwo$. Despite the large uncertainties, this 
observation complements the measurement of a large suppression of D meson production at high $\ptrans$ (about a factor 5 at $\ptrans$$\sim$10~$\gev/c$) in
central Pb--Pb collisions, which constitutes an evidence of in-medium charm energy loss and 
points to a strong interaction of charm with the QGP. The comparison of the $\RAA$ of D mesons and of 
secondary $\jpsi$ from b-hadron decays suggests that beauty loses less energy than charm, corroborating the prediction
of a dependence of energy loss from the quark mass induced by the dead-cone effect. However, a satisfactory description
of the interaction of charm and beauty quarks with the medium is still missing as highlighted by the absence of a model able to reproduce
simultaneously the $\RAA$ and $\vtwo$ of D mesons and electrons from heavy-flavour hadron decay. 
Although the results obtained at the LHC after the first 3 years of data taking can be casted into a qualitatively understood frame, 
more precise and ``differential'' measurements are needed to focus the emerging picture of
heavy-flavour production and interaction in the QGP. The data collected in the next runs and, even more, the upgrade of the detectors, during
the long shut down in 2018, should allow to reach this target and bring experimental heavy-ion physics into the charm and beauty era
of QGP.

%

\begin{thebibliography}{}
%
%
\bibitem{pbmJSNature} P. Braun-Munzinger and J. Stachel, Nature {\bf 448}, 302-309 (2007).
\bibitem{gyulassy} M. Gyulassy and M. Plumer, Phys. Lett. {\bf B243}, 432 (1990).
\bibitem{bdmps} R. Baier, Y.~L. Dokshitzer, A.~H. Mueller, S. Peigne and D. Schiff, 
Nucl. Phys. {\bf B484}, 265 (1997).
\bibitem{thoma}
M. H. Thoma and M. Gyulassy, Nucl. Phys. {\bf B351}, 491 (1991).\\
E. Braaten and M. H. Thoma, Phys. Rev. {\bf D44}, 1298 (1991); Phys. Rev. {\bf D44}, 2625 (1991).
\bibitem{deadcone} Y.~L.~Dokshitzer, D.~E.~Kharzeev,  
  Phys.\ Lett.\  {\bf B519 }, 199 (2001). \\
  N.~Armesto, C.~A.~Salgado and U.~A.~Wiedemann,  Phys.\ Rev.\  {\bf D69}, 114003 (2004).\\
  M.~Djordjevic, M.~Gyulassy, Nucl. Phys. {\bf A733} 265 (2004).\\
  B.-W. Zhang, E. Wang and X.-N. Wang, Phys. Rev. Lett. {\bf 93}, 072301 (2004).\\
  S.~Wicks, W.~Horowitz, M.~Djordjevic and M.~Gyulassy,
  Nucl.\ Phys.\  {\bf A783 }, 493 (2007).
\bibitem{Armesto:2005iq} N.~Armesto, A.~Dainese, C.A.~Salgado and U.A.~Wiedemann,
  Phys. Rev. {\bf D71}, 054027 (2005).
\bibitem{glauber} R.~J.~Glauber in Lectures in Theoretical Physics, NY, {\bf Vol. 1}, 315 (1959).\\
  M.~Miller {\it et al.}, Ann.\ Rev.\ Nucl.\ Part.\ Sci.\ {\bf 57}, 205 (2007).
\bibitem{vtwoOllitrault} J. Y. Ollitrault, Phys. Rev. {\bf D46}, 229 (1992).
\bibitem{bamps} O. Fochler, J. Uphoff, Z. Xu and C. Greiner, J. Phys {\bf G38}, 124152 (2011).
\bibitem{HFurQMD} T. Lang, H. van Hees, J. Steinheimer and M. Bleicher, arXiv:1211.6912 [hep-ph]. \\
T. Lang H. van Hees, J. Steinheimer, Y.-P. Yan and M. Bleicher, arXiv:1212.0696 [hep-ph]. 
\bibitem{CollLPMrad} P. B. Gossiaux, J. Aichelin and R. Bierkhandt, Phys. Rev {\bf C79}, 044906 (2009).  \\
 P. B. Gossiaux, J. Aichelin, T. Gousset and V. Guiho, J. Phys {\bf G37}, 094019 (2010).
\bibitem{powlang} W. M. Alberico et al., Eur. Phys. J. {\bf C71}, 1666 (2011).\\
W. M. Alberico et al., J. Phys. {\bf G38}, 164144 (2011).
\bibitem{whdg} W. A. Horowitz and M. Gyulassy, J. Phys. {\bf G38}, 124152 (2011).
\bibitem{TamuRappEtAl} M. He, R. J. Fries and R. Rapp, Phys. Rev. {\bf C86}, 014903 (2012).
\bibitem{jpsiSatz} T. Matsui and H. Satz, Phys. Lett. {\bf B178}, 416 (1986).
\bibitem{jpsiSequential} S. Digal, P. Petreczky, H. Satz, Phys. Rev.~{\bf D64}, 0940150 (2001).\\
  A. Mocsy, P. Petreczky,  Phys. Rev. Lett. {\bf 99}, 211602 (2007)
\bibitem{jpsiSPS} B. Alessandro et al. (NA50 Collaboration), Eur. Phys. J. {\bf C39}, 335 (2005).
\bibitem{jpsiPHENIX} A. Adare et al. (PHENIX Collaboration), Phys. Rev. Lett. {\bf 98}, 232301 (2007). \\ 
  A. Adare et al. (PHENIX Collaboration), Phys. Rev. {\bf C84}, 054912 (2011).
\bibitem{jpsiSHM} P. Braun-Munzinger and J. Stachel, Phys. Lett. {\bf B490}, 196 (2000).
\bibitem{jpsiAndronic} A. Andronic, P. Braun-Munzinger, K. Redlich, and J. Stachel, J. Phys. {\bf G38}, 124081 (2011).
\bibitem{jpsiZhaoRapp} X. Zhao and R. Rapp, Nucl. Phys. {\bf A859}, 114 (2011).
\bibitem{jpsiLiuQuXu} Y.-P. Liu, Z. Qu, N. Xu, and P.-F. Zhuang, Phys.Lett. {\bf B678}, 72 (2009).
\bibitem{pPbPredictions} J. L. Albacete et al., Int. J. Mod. Phys. {\bf E Vol. 22} 1330007 (2013), arXiv:1301.3395 [hep-ph].
\bibitem{fonll} M. Cacciari, M. Greco and P. Nason, JHEP {\bf 9805}, 007 (1998).
\bibitem{Fasel} M. Fasel, these proceedings.
\bibitem{CacciariThisProc} M. Cacciari, these proceedings.
\bibitem{Woehri} H. Woehri, these proceedings.
\bibitem{ALICEjpsiPaper} B.~Abelev~et~al.~(ALICE Collaboration), Phys. Rev. Lett. {\bf 109}, 072301 (2012).
\bibitem{jpsiFerreiro} E. G. Ferreiro, arXiv:1210.3209 [hep-ph]; private communication.
\bibitem{Manceau} L. H. A. Manceau, these proceedings.
\bibitem{ATLASjpsiRcp} G. Aad et al. (ATLAS Collaboration), Phys.Lett. {\bf B697}, 294-312 (2011).
\bibitem{CMSjpsiPaper} S. Chatrchyan et al. (CMS Collaboration), JHEP {\bf 05}, 063 (2012).
\bibitem{CMSjpsiPAS} CMS Collaboration, CMS PAS HIN-12-014 (2012).
\bibitem{epsZeroNine} K. J. Eskola, H. Paukkunen and C. A. Salgado, JHEP {\bf 0904}, 065 (2009).
\bibitem{pPbshadowingVogt} R. Vogt, Int. J. Mod. Phys. {\bf E Vol. 22}, 1330007 (2013), arXiv:1301.3395 [hep-ph]; private communication.
\bibitem{jpsipPbArleo} F. Arleo and S. Peign, arXiv:1212.0434 (2013).\\
  F. A. et al., arXiv:1304.0901 (2013).
\bibitem{cgcFuji} K. W. Hirotsugu Fujii, arXiv:1304.2221 (2013).
\bibitem{jpsiVtwo} E. Abbas et al. (ALICE Collaboration) arXiv:1303.5880 [nucl-ex].
\bibitem{psiTwosCMS} CMS Collaboration, CMS PAS HIN-12-007.
\bibitem{jpsiPsiTwosArnaldiQM} R. Arnaldi for the ALICE Collaboration, arXiv:1211.2578 [nucl-ex].
\bibitem{CMSupsilon} S. Chatrchyan et al. (CMS Collaboration), Phys. Rev. Lett. {\bf 109}, 222301 (2012).
\bibitem{aliceDmesons} B. Abelev et al. (ALICE Collaboration), JHEP {\bf 09}, 112 (2012). 
\bibitem{zaidaQM} Z. Conesa del Valle for the ALICE Collaboration, arXiv:1212.0385 [nucl-ex].
\bibitem{Bianchin} C. Bianchin, these proceedings.
\bibitem{muonPRL} B. Abelev et al. (ALICE Collaboration) Phys. Rev. Lett. {\bf 109}, 112301 (2012).
\bibitem{ATLASmuonRcp} ATLAS Collaboration, ATLAS-CONF-2012-050 (2012).
\bibitem{innocentiDs} G. M. Innocenti for the ALICE Collaboration, arXiv:1210.6388 [nucl-ex].
\bibitem{DsRaphelsky} I. Kuznetsova and J. Rafelski, J. Phys. {\bf G32} S499-S504 (2006).
\bibitem{Dmesonv2} B. Abelev et al. (ALICE Collaboration), arXiv:1305.2707 [nucl-ex].
\bibitem{VoloshinV2} A. M. Poskanzer and S. Voloshin, Phys.Rev. {\bf C58}, 1671 (1998), arXiv:nucl-ex/9805001 [nucl-ex].
\bibitem{hadronV2} B. Abelev et al. (ALICE Collaboration), arXiv:1205.5761 [nucl-ex].
\end{thebibliography}
%
%

\end{document}